\documentclass[aps,twocolumn,groupedaddress,floatfix]{revtex4-1}

\usepackage{amssymb}
\usepackage{graphicx}

\begin{document}
\title{Application of lasers to ultracold atoms and molecules}

\author{H\'el\`ene Perrin$^{1,6}$}
\author{Pierre Lemonde$^{2,6}$}
\author{Franck Pereira dos Santos$^{2,6}$}
\author{Vincent Josse$^{3,6}$}
\author{Bruno Laburthe Tolra$^{1,6}$}
\author{Fr\'ed\'eric Chevy$^{4,6}$}
\author{Daniel Comparat$^{5,6}$}

\affiliation{$^1$Laboratoire de physique des lasers, CNRS and Universit\'e Paris 13, 99 avenue J.-B. Cl\'ement, 93430 Villetaneuse}
\affiliation{$^2$LNE-SYRTE, Observatoire de Paris, CNRS and UPMC, 61 Avenue de l'Observatoire, 75014 Paris, France}
\affiliation{$^3$LCFIO, Institut d'Optique Graduate School and CNRS, Campus Polytechnique, RD128, 91127 Palaiseau Cedex}
\affiliation{$^4$Laboratoire Kastler Brossel, ENS, CNRS, UPMC, 24 rue Lhomond, 75005 Paris, France}
\affiliation{$^5$Laboratoire Aim\'e Cotton, CNRS, Univ Paris-Sud, B\^at. 505, 91405 Orsay, France}
\affiliation{$^6$IFRAF, Institut Francilien de Recherche sur les Atomes Froids, 45 rue d'Ulm, 75005 Paris, France}

\begin{abstract}
In this review, we discuss the impact of the development of lasers on ultracold atoms and molecules and their applications. After a brief historical review of laser cooling and Bose-Einstein condensation, we present important applications of ultra cold atoms, including time and frequency metrology, atom interferometry and inertial sensors, atom lasers, simulation of condensed matter systems, production and study of strongly correlated systems, and production of ultracold molecules.
\end{abstract}
\maketitle

\section{Introduction}
\label{intro}

The manipulation of dilute gases with light is a good example of a field where a major technological step -- the development of lasers -- triggered important achievements in fundamental research -- high resolution spectroscopy, laser cooling and trapping, and Bose-Einstein condensation -- which in turn contributed to the development or improvement of new technologies -- atomic fountain clocks, atomic gyrometers, atom lasers, quantum simulators... The mechanical effect of light on matter was first demonstrated in 1901, where radiation pressure was evidenced independently by Lebedev, and Nichols and Hull~\cite{LebedevNichols1901}. Otto Frisch demonstrated in 1933 that individual particles from a beam of sodium are deflected by the radiation pressure of a sodium lamp~\cite{Frisch1933}. But these early demonstration experiments were not developed further until the advent of lasers. Only ten years after the first demonstration of the laser, Ashkin proposed to use laser light to manipulate the external degrees of freedom of atoms~\cite{Ashkin1970}. The first experiments using lasers for deflecting an atomic beam were performed two years later~\cite{PicqueSchieder1972}.

The mechanical action of light on matter, described as light forces, is linked to the momentum transfer of a photon recoil $\hbar k$ in the absorption or emission process of one single photon of wave vector $k$. Light is coupled to atoms by the interaction between the electric light field and the atomic dipole. Two kinds of light forces can be distinguished, radiation pressure and the dipole force\,\cite{CohenHouches1990}. Radiation pressure is a \textit{dissipative} force and corresponds to the absorption of a photon from a light source followed by the spontaneous emission of another photon. As spontaneous emission is a random process with equal probabilities in opposite directions, a net average force is built in the direction of the light wave vector when this process is repeated. Radiation pressure is particularly efficient with a laser tuned on a strong cycling transition of an atom or ion, and accelerations of order $10^5$\,m$\cdot$s$^{-2}$ can be reached with moderate laser powers of a few mW. Radiation pressure depends on the laser frequency like the photon scattering rate and is maximum on resonance. On the other hand, a redistribution of photons in the light field occurs when the absorption is followed by a \emph{stimulated} emission. These processes lead to an energy shift of the atomic states known as light shift. The variation of the light shift in space, linked to intensity gradients in the light field, is responsible for the dipole force, which is \textit{conservative}. The dipole force vanishes on resonance and is opposite for opposite detunings of the light frequency with respect to the atomic transition. Depending on the value of the detuning, either one or the other force can be dominant. Historically, the first experiments on atomic manipulation made use of the radiation pressure only as the dipole force becomes important for rather cold atomic samples.

In 1975, H\"ansch and Schawlow~\cite{Haensch1975} suggested to take benefit of the Doppler effect to make the radiation pressure velocity dependent: laser cooling was born. At the same time, a similar idea was proposed by Wineland and Dehmelt to cool trapped ions~\cite{Wineland1975}. Tuning two counter-propagating lasers below the atomic resonance favours the absorption from the laser propagating against the atom, which is set closer to resonance by the Doppler shift. Hence, the radiation pressures from the two lasers are unbalanced and a net force acts against the atomic velocity, leading to dissipation and cooling. This Doppler cooling scheme can be generalised to all three dimensions of space with six laser beams. It leads to very low but finite velocities, limited to typically a few cm/s by the fluctuation of the instantaneous force due to absorption and spontaneous emission of individual photons. As the energy extracted at each absorption-emission cycle, the recoil energy, is rather low, many cycles are necessary to efficiently cool thermal atoms. Cooling therefore requires an almost closed transition, which can be found in many atoms. However, this requirement is very demanding in the case of molecules.

If laser cooling could be implemented in 1978 soon after these proposals with trapped ions~\cite{Wineland1978}, it was not the case for neutral atoms. Indeed, due to a smaller interaction of neutral atoms with external fields, conservative traps are not deep enough to first trap atoms at room temperature before applying laser cooling, as is the case for ions, and the interaction time between lasers and thermal atomic beams is not large enough to allow for efficient cooling. An additional decelerating step was necessary before laser cooling could be applied, and this was achieved in 1985 with the first Zeeman slower~\cite{Prodan1985}: a laser propagating against an atomic beam is maintained into resonance during the deceleration process thanks to an inhomogeneous magnetic field, the Zeeman shift compensating for the Doppler shift everywhere on the atomic trajectory. The first implementation of laser cooling in a six-beam molasses, loaded by a Zeeman slower, followed immediately~\cite{Chu1985}.

An important step toward the applications of cold atoms was the implementation of the first magneto-optical trap by Raab \emph{et al.} in 1987, following a suggestion of Dalibard~\cite{Raab1987}. The basic idea is to make the radiation not only velocity- but also position-dependent by the addition of a magnetic field gradient to the setup. The Zeeman shift splits the magnetic substates of both the ground state and the excited state, which makes the transition frequency sensitive to the local magnetic field and hence to the position. This results in a restoring force. The magneto-optical trap setup was even simplified in 1990 by Monroe \emph{et al.} who demonstrated that the atoms could be loaded directly from the low velocity tail of a vapour at room temperature~\cite{Monroe1990}. The magneto-optical trap represented a major step in the quest for large phase space densities in atomic vapours~\cite{Townsend1995}, essential for Bose-Einstein condensation.

Laser cooling has proven to be an extremely powerful technique to reach low temperatures in an atomic vapour. Indeed, the temperature of $40\,\mu$K obtained with sodium atoms was even lower than the $240\,\mu$K predicted by a theory based on Doppler cooling only~\cite{Lett1988}. This is due to the combination of optical pumping and differential light shift between magnetic substates of the atomic ground state, atoms being always pumped into the substate with the lowest energy~\cite{Dalibard1989}. The demonstration of laser cooling of atoms and the theoretical developments for its understanding were rewarded by the 1997 Nobel prize in physics~\cite{Nobel1997}.

With the progress of laser cooling, the use of the dipole force to act on atoms became relevant. With blue-detuned lasers -- detuned above the atomic transition -- atoms are pushed away from the regions of high intensity, whereas they are attracted towards the regions of high intensity in the case of a red detuning. The dipole force is proportional to the gradient of the light intensity, and Cook and Hill proposed in 1982 to use the huge intensity gradient of an evanescent wave at the surface of a dielectric to repel atoms from the surface~\cite{Cook1982}. This kind of \emph{atom mirror} was demonstrated by Balykin \emph{et al.} a few years later~\cite{Balykin1987}. Ashkin proposed in 1978 to trap atoms at the focus point of a red-detuned laser~\cite{Ashkin1978}. This was realised in 1986 when Chu \emph{et al.} loaded atoms from an optical molasses into the first \emph{dipole trap}~\cite{Chu1986}. Since then, dipole traps have became an important tool of ultracold atom experiments~\cite{Grimm2000}. In particular, optical lattices~\cite{Grynberg2001}, where atoms are trapped in the nodes or antinodes of a light standing wave created by the interference of laser beams, have had important applications in quantum simulation~\cite{Feynman1982,Bloch2008,Simon2011} as well as in time metrology, as described in section~\ref{Lemonde}.

As a matter of fact, the importance of laser cooling for time and frequency metrology was recognised very early. The long measurement times available with very slow atoms made possible in 1989 the successful realisation of the Zacharias fountain~\cite{Kasevich1989}. The idea is to launch atoms vertically to let them interact twice with the microwave cavity tuned to the caesium hyperfine frequency defining the second, and detect Ramsey fringes. The delay of typically 0.5\,s between the two interactions sets the uncertainty on the measurement. The atom fountain is used now on an everyday basis to determine the International Atomic Time (TAI). The recent development of optical clocks with neutral atoms in optical lattices was again allowed by laser cooling and laser trapping in standing waves, see section~\ref{Lemonde}. Here, a transition in the optical domain is used, improving greatly the uncertainty in the relative frequency. Moreover, the advent of femtosecond combs, celebrated by the 2005 Nobel prize~\cite{Nobel2005}, made possible the direct comparison of this optical frequency with the microwave frequency standard. 

Cold atoms are also used in atom interferometers as very sensitive inertial sensors, as discussed in section~\ref{Pereira}: best atomic gravimeters and gyrometers compete with state of the art instruments~\cite{Cronin2009}. Here, the coherence of the atomic source will be an advantage. As monomode lasers of large coherence length greatly improved light interferometry, the development of an \emph{atom laser} is of great interest for improving the coherence of atom interferometers. A giant step towards this goal has been made in 1995 with the first observation of Bose-Einstein condensation in dilute gases, for which the Nobel prize was attributed in 2001 to Cornell, Ketterle and Wieman~\cite{Nobel2001}. This was allowed by laser cooling followed by evaporative cooling in a magnetic trap~\cite{Anderson1995}. An atom laser, as described in more detail in section~\ref{Josse}, is a coherent atomic source, in which all the atoms occupy a single quantum mode. Indeed, a Bose-Einstein condensate (BEC) fulfils this criterion: below the critical temperature, there is a macroscopic population of the ground state, a single quantum state.

The coherence of the BEC considered as a new source for atom optics was demonstrated soon after its first observation~\cite{Andrews1997}. Even if the condensation is obtained with dilute gases, interactions between atoms play an essential role in the physics of the BEC. This can lead to phase diffusion and degrade the coherence length, especially in 1D atomic guides~\cite{Bloch2008RMP}. On the other hand, interactions provide non linearity which may improve interferometers through, \textit{e.g.}, squeezing, in analogy to non linear quantum optics\,\cite{Gross2010}. Interactions are also responsible for the superfluidity of the condensate, evidenced by the presence of quantum vortices when the gas is set into rotation\cite{Madison2000}, in analogy with vortex lattices in superconductors. In fact, quantum gases can be seen as quantum liquids, which bridges atomic physics and condensed matter physics. Two striking examples are the observation in 2002 of the Mott insulator to superfluid transition with atoms in an optical lattice~\cite{Greiner2002}, an analogue to the Mott transition in solids, and more recently, the observation of Anderson localisation with cold atoms in a one-dimensional system~\cite{Billy2008,Roati2008} described in section~\ref{Josse}.

More generally, degenerate quantum gases can be considered as model systems, with easily tunable parameters, for difficult problems in condensed matter physics. In particular, the use of dipole traps -- atom traps relying on the dipole force -- to confine and produce BECs~\cite{Barrett2001} opened the way to new experiments using a magnetic field as a free parameter: the study of new quantum phases for multi-component BECs\,\cite{Ho1998,Machida1998}, or the tuning of atomic interactions and non linearity through a Feshbach resonance~\cite{Inouye1998} to investigate strongly interacting regimes. Examples of the use of dipole traps for the production and the study of degenerate gases are given in section~\ref{Laburthe}. Fermions can also be laser cooled and evaporatively cooled to degeneracy, and the analogy with electrons in condensed matter is even more natural in this case. The regime of strongly interacting fermions is now available with Feshbach resonances in dipole traps, see section~\ref{Laburthe}.

Finally, the physics of cold matter has also been developed with more complex systems. Cold molecules can be formed from ultracold atoms through a Feshbach resonance and even be brought to Bose-Einstein condensation~\cite{Herbig2003}. The grail of producing cold molecules of astronomical, biological or fundamental interest for high resolution spectroscopy or controlled cold chemistry is not reached yet. However, a major progress towards laser cooling of molecules has been accomplished recently both with neutral molecules~\cite{Viteau2008,Shuman2010} and molecular ions~\cite{Staanum2010,Schneider2010}, and the expected developments are presented in section~\ref{Comparat}.

\section{Optical lattice clocks}
\label{Lemonde}

Lasers are at the heart of the recent developments in the field of atomic clocks. A first revolution happened around 1990 with the advent of clocks using laser cooled atoms, known as atomic fountains\,\cite{Kasevich1989,Clairon1991}. The clock transition of Cs or Rb atoms cooled down to about $1\,\mu$K can be probed for up to a second, i.e. two orders of magnitude longer than in traditional atomic beam apparatus. This leads to proportionally narrower atomic resonances. In addition, with these slow atoms, key physical effects like the Doppler frequency shift which is a long standing limitation to the clock accuracy, are strongly reduced. This results in a dramatic improvement of the clock performance and atomic fountains now come close their ultimate limits. The best devices exhibit a quantum limited residual frequency noise close to $10^{-14}\,\tau^{-1/2}$ with $\tau$ the averaging time in seconds and a control of systematic effects at a level of $3\times 10^{-16}$ in relative frequency\,\cite{Bize2005,Bauch2006}.

In atomic fountains the frequency reference is a hyperfine transition of the atomic ground state at a frequency of 9.2 GHz for Cs and 6.8 GHz for Rb. It has long been anticipated that another revolution would be possible by switching to a transition in the optical domain, at a frequency that is 4 orders of magnitude larger\,\cite{Hollberg2005}. Indeed both the quantum limit to frequency noise and the relative magnitude of most frequency shifts scale as $1/\nu$ with $\nu$ the clock frequency. This however required solving two major issues. The first one is to be able to effectively measure optical frequencies and compare optical clocks which operate at a much too high frequency for the electronic devices used for that purpose in the microwave and radio-frequency (RF) domains. The solution to this problem came in the late 90's with the advent of femtosecond frequency combs, another revolution which boosted the development of optical frequency standards since then\,\cite{Diddams2000,Nobel2005}.

The second issue is to tame the Doppler effect which remains the dominant term in the accuracy budget of fountains and is the major exception to the $1/\nu$ rule stated above. It is well known that confining the reference particles to a region of space that is smaller than the transition wavelength (sub-micrometer for an optical transition) is a way to cancel motional effects\,\cite{Dicke1953,Vanier1989}. This is the Lamb-Dicke regime which is a key ingredient for narrow resonances in hydrogen masers or M\"{o}ssbauer resonances. Confinement however generally shifts the clock frequency and it was thought for a long time that this idea could only be applicable for trapped ions in high accuracy clocks. Thanks to their external charge, ions can indeed be confined in relatively low fields. It was shown recently that all systematic effects could be controlled down to better than $10^{-17}$ in a clock using a single trapped Al$^+$ ion\,\cite{Chou2010a}. This clock sets a new state of the art in the field and outperforms atomic fountains by more than an order of magnitude. Using trapped neutral atoms in an optical clock still remains highly desirable since a large number of particles can be trapped and interrogated simultaneously thanks to the limited interaction between neutrals. The quantum limit to the detection signal to noise ratio scales as $\sqrt{N}$, with $N$ the number of contributing particles so that an optical clock with $10^4$ to $10^6$ atoms could potentially surpass its single ion counterpart by 2 to 3 orders of magnitude in terms of residual frequency noise. 
Trapping neutral atoms in the Lamb-Dicke regime was experimentally demonstrated in the early 90's by using optical lattices\,\cite{Grynberg2001}. Optical lattices are periodic potentials formed by the interference of several laser beams which generate a set of trapping wells at a sub-micrometer scale. Efficient trapping however requires large fields which at first sight cannot be controlled at the required level. Beating gravity in an optical lattice typically demands a shift of the energy levels of tens of kHz, i.e. $10^{-10}$ of an optical frequency\,\cite{Lemonde2005}. 
This implies a control of the lattice depth at the $10^{-8}$ level if one aims at a fractional accuracy in the $10^{-18}$ range !
 
This frequency shift problem can however be circumvented in an optical lattice clock thanks to a smart tailoring of the light shifts, as proposed in 2001 by H. Katori\,\cite{Katori2002,Katori2003}. Katori proposed
a configuration where the shift of both clock states exactly match. An optical lattice is based on the dipole force which generally speaking depends on the trap laser polarisation, intensity and frequency\,\cite{Grimm2000}. The clock transition in an optical lattice clock couples two states with zero total electronic angular momentum ($J=0$). These spherically symmetrical states experience a dipole force that is to leading order independent of the field polarisation. In addition, the coupling between these states is extremely weak. It is in fact forbidden to all orders for a single photon coupling, so that the excited state is metastable allowing for long coherence times and consequently narrow resonances (the same kind of transition is used in an Al$^+$ or In$^+$ clock for this reason). For $^{87}$Sr, which was the first atom for which the scheme was proposed, the transition is weakly allowed by hyperfine coupling only. Its natural linewidth is 1\,mHz and will most probably never constitute a limit to the clock performance\,\cite{Courtillot2003}. The key ingredient of the lattice clock scheme is to tune the lattice frequency to the so-called \textit{magic frequency} such that the intensity dependence of the dipole force exerted on the clock states are identical to leading order. The only remaining critical parameter of the trapping field is in principle its frequency which can be controlled to its magic value extremely well... ultimately by an atomic clock! Several atoms have the requested energy level structure for an optical lattice clock. This is the case for most alkaline-earth (Sr, Ca, Mg) and related atoms (Yb, Hg,...) and several projects using these atoms have been started since 2001 mainly with Sr and Yb\,\cite{Takamoto2005,Ludlow2006,Letargat2006,Barber2006,Hachisu2008,Petersen2008,Lisdat2009}.

The proposal developed by Katori is based on leading order arguments, while several higher order effects have been identified as potential issues to the ultimate performance and should definitely be considered when the requested light shift cancelation lies in the $10^{-8}$ range. In addition these effects are usually difficult to predict theoretically with the requested accuracy. Several experiments progressively demonstrated that these potential issues actually do not constitute a serious limitation to the clock performance, down to the $10^{-17}$ level for Sr and Yb\,\cite{Brusch2006,Boyd2007a,Barber2008}. 

The most advanced optical lattice clock now have an accuracy in the low $10^{-16}$ range\,\cite{Ludlow2008,Lemke2009}, already better than atomic fountains and approaching the best ion clocks performance. This corresponds to a control of the frequency of the laser probing the atoms to within less than $0.1\,$Hz. In addition, several measurements of the Sr clock transition frequency performed in different laboratories in totally different setups display a perfect agreement, strengthening further the confidence in the potential of these new apparatuses\,\cite{Blatt2008}. And much room for improvement remains. In terms of residual frequency noise, the state-of-the-art is marginally better than single ion clocks and approaches $10^{-15}\,\tau^{-1/2}$. This is about two orders of magnitude away from the expected quantum limit in these clocks using a large number of atoms, and results from a technical though very difficult issue, associated to the residual frequency noise of the laser probing the clock transition\,\cite{Quessada2003}. Several schemes have been proposed to either dwarf the measurement sensitivity to this noise source and/or reduce the noise source itself\,\cite{Westergaard2010,Meiser2009,Lodewyck2010}. It is doubtless that in the coming years at least an order of magnitude can be gained which would allow extending the coherence time of the clock laser to tens of seconds or more by stabilisation to the atomic reference.

Optical lattice clocks are a topical example of the use of lasers in physics. Laser cooled atoms confined in a laser trap are excited by an ultra-stable probe laser which in turns allows to further stabilise and accurately control the long term properties of this laser frequency. Frequency-combs based on femtosecond laser pulses are then used to transfer these properties to other regions of the electromagnetic spectrum where they can be used to either generate a time-scale or perform more physics experiments\,\cite{Udem2009}. The development of lattice clocks is indeed interesting in itself for the progress of the field of time and frequency metrology. But other very promising applications in other fields of physics are anticipated and already under investigation. An interesting case is the study of collisions between cold atoms in the lattice. The collisional properties of confined quantum gases is definitely a hot topic and the information retrieved from clock experiments are complementary to the ones deduced from the dynamics of degenerate quantum gases\,\cite{Gibble2009,Campbell2009,Blatt2009}.
In a totally different field, optical lattice clocks can be used for fundamental physics experiments. The comparison and gathering of the measurements performed in different institutes allowed setting a very stringent constrain on the coupling of gravity with other fundamental interactions by looking the Sr/Cs clock frequency ratio in the various gravitational potentials provided by the ellipticity of the Earth rotation around the Sun\,\cite{Blatt2008}. In the future, more information could be gathered by sending these clocks in space in Earth-orbit or in deep space and several such projects have already been proposed to the European Space Agency (ESA) and are under investigation\,\cite{SchillerWolf2009}.

\section{Atom interferometry}
\label{Pereira}
Another important application of laser cooled atoms concerns the development of atom interferometers. Because of the wave nature of matter, interferometry phenomena, which any physicist is familiar with in the case of light, can also be observed with particles, atoms or molecules. To reveal this fascinating feature, a major difficulty is to develop tools to manipulate matter waves, whose associated de Broglie wavelengths are in general relatively small compared to the case of light. For example, de Broglie wavelengths for thermal atomic beams lie in the tens of picometre range. In this respect, (ultra)-cold atoms are of great interest for interferometry experiments because of their comparatively much larger de Broglie wavelengths, typically in the micrometer range.
Early experiments that demonstrated this wave nature, mainly through diffraction properties, were realised first with elementary particles and later with atoms. Material objects such as crystals \cite{Estermann1930}, or later micro-fabricated slits \cite{Carnal1991} or gratings were used to diffract matter waves and create interferometers \cite{Keith1988}. Over all the various tools that have been demonstrated for the manipulation of matter waves, lasers have quickly proven to be the most efficient, because of their intrinsic properties (monochromaticity, brightness, accurate knowledge of their wavelengths...).

The key feature of the interaction of matter waves with lasers lies in the exchange of momentum between the atoms or molecules and the light field. Consider a transition in an atom between two long-lived internal states, for instance a ground state $|g\rangle$ and a metastable excited state $|e\rangle$, that can be excited with lasers, through a one- or several-photon transition. The interaction with the laser field leads to a Rabi oscillation, which leaves the atom in a quantum superposition of ground and excited states when the laser is switched off. Due to conservation of momentum, an initial state $|g,p\rangle$ couples to $|e,p+\hbar k\rangle$, where $p$ is the initial momentum of the atom and $\hbar k$ is the momentum of the photon. These two coupled states differ not only in internal state, but also in momentum state. As a consequence, a laser pulse puts an initial wave packet into a superposition of two partial wavepackets that separate when evolving freely after the interaction with the laser. Pulses of adjusted intensity and/or duration allow to control the relative weights of the two partial wavepackets in this superposition. The particular cases of $\pi/2$ and $\pi$ pulses, which correspond to equal weights or full transfer, are respectively the analogue of the beamsplitters and the mirrors for the matter waves.

With such a tool, a Mach Zehnder type interferometer can be created using a sequence of three pulses, $\pi/2-\pi-\pi/2$ that allow to split, redirect and recombine the wavepackets. The atomic state at the output of the interferometer then depends on the difference of the phases accumulated by the atoms along the two paths. This interferometer geometry is used by most of the atomic inertial sensors \cite{Peters2001,Gustafson2000}. In these instruments, the interferometer phase, which scales with the square of the interferometer duration, depends on the acceleration along the direction of the lasers and of the rotation rate along a direction perpendicular to the area enclosed by the two arms of the interferometer. For a duration on the order of 100 ms, which can easily be obtained with cold atoms, the phase shift is extremely large (hundreds of thousands of radians for an acceleration of the order of g, and a few radians for a rotation rate as small as the Earth's rotation rate). In most cases, two photon Raman transitions are used that couple the two hyperfine ground states in an alkali atom \cite{Kasevich1991}. The sensitivity to inertial forces arises from the interaction with the Raman lasers, as at each pulse the difference between the phases of the Raman lasers gets imprinted on the diffracted wavepacket. The final phase shift finally depends on the positions of the atoms, modified by inertial forces, with respect to the planes of equal phase difference between the lasers, which act as a precise ruler linked to the laboratory frame. The exact knowledge of the wavelength of the lasers thus allows realising absolute measurements of acceleration and rotation rates. Atomic inertial sensors have already reached performances comparable with state of the art instruments \cite{Merlet2010}, and can still be improved. The technology is now mature enough for industrial developments. These instruments have a wide range of applications, from navigation to geophysics but also in fundamental metrology (the gravimeter described in \cite{LeGouet2008} is being developed within the French watt balance project, which aims at redefining the kilogram through the measurement of the Planck constant \cite{Geneves2005}).

Many other interferometer geometries have been demonstrated, based on different pulse sequences \cite{Borde1989,Riehle1991,Canuel2006}, or on different ``beamsplitters'', such as standing waves, which allow for Bragg \cite{Rasel1995,Giltner1995} or Kapitza Dirac \cite{Cahn1997} diffraction. The sensitivity of the interferometer can be improved by using larger momentum transfer in the beam splitting process, based for instance on double diffraction \cite{Leveque2009}, Bloch oscillations \cite{Clade2009,Muller2009} or multi-photon transitions \cite{Weitz1994,Muller2008}.

Atom interferometers based on trapped or guided geometries would allow reaching large interrogation times and thus excellent sensitivities without the size constraints of having the atoms in free fall in the interferometer. This would thus allow for a drastic reduction in the size of the apparatuses with respect to the experiments described above. A drawback though lies in the perturbations inherent to the propagation in the guide or to the trap, which need to be perfectly controlled, and to the limited coherence due to interatomic interactions in these systems \cite{Grond2010}. Several demonstrations have nevertheless been made \cite{Shin2004,Wang2005,Garcia2006,Schumm2005b,Albiez2005,Su2010}, which might open the way for the development of a new class of instruments.

Finally, since they allow for absolute measurements, laser based atom interferometers are also excellent tools for fundamental physics. They are used for instance for accurate measurements of the fine structure constant \cite{Bouchendira2011}, the Newtonian gravitational constant $G$ \cite{Fixler2007,Lamporesi2008}, and of atomic polarisabilities \cite{Ekstrom1995}. Proposals have been made for measurements of the Lense-Thirring effect \cite{Jentsch2004}, tests of Newtonian gravity either at short range \cite{Wolf2007} or at the scale of the solar system \cite{Wolf2009}, tests of the equivalence principle \cite{Dimopoulos2007,Varoquaux2009,Gaaloul2010}, detection of gravitational waves \cite{Dimopoulos2008}, or tests of atom neutrality \cite{Arvanitaki2008}.

\section{Atom lasers}
\label{Josse}

Bose-Einstein condensation corresponds to a macroscopic accumulation of particles in the ground state of the system. All atoms being in the same mode, i.e. having the same wavefunction, the BEC is the matterwave analogue of a photon laser field inside an optical cavity. In analogy with a propagating laser field, a so-called \textit{atom laser} can be formed by coherently extracting atoms from the BEC. Not surprisingly, such atomic outcouplers, which play the role of partially transmitting mirrors in optics, have been quickly developed after the achievement of Bose-Einstein condensation in dilute gases. In 1997, in the group of W. Ketterle at MIT, a pulsed RF field was used to perform a transition between the BEC state (a low-field seeker Zeeman sublevel trapped in a magnetic field) and a nearly magnetic insensitive (untrapped) state, leading to atomic wave packets falling under gravity: the first atom laser\,\cite{Mewes:1997}. 

Other atom laser prototypes followed shortly after. In Munich, a quasi-continuous atom laser beam was produced using a similar RF outcoupler\,\cite{Bloch:1999}, but taking advantage of a new, ultra-stable magnetic field configuration. Meanwhile, another kind of atom laser was produced at Yale from the Landau-Zener decay of a BEC loaded in a vertical optical lattice: there the constructive interference between the decay occurring at different lattice sites led to a pulsed emission of a coherent matterwave, in close analogy with mode-locked photonic lasers\,\cite{Anderson:1998}. Last, a two-photon Raman transition outcoupler, transferring a large momentum kick to the atoms, was realised at NIST to generate a well collimated and directional atomic beam\,\cite{Hagley:1999}.

\begin{figure}[t!]
\centering
\includegraphics[width=\linewidth]{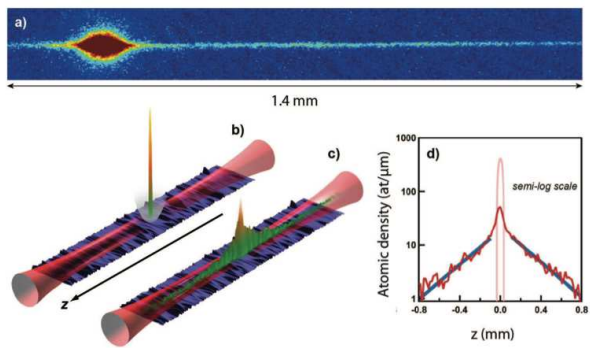}
\caption{Top (a): Image by absorption of an atom laser directly outcoupled in an horizontal optical waveguide (from ref.\,\cite{nous:GAL}). Bottom: Experimental scheme of~\cite{Billy2008} for the demonstration of Anderson Localisation with ultracold atoms. Following the proposal~\cite{LSP:2009}, a somewhat crude version of an atom laser (the whole initial BEC (b) being released) is allowed to expand along a guide and is stopped in presence of a laser speckle disorder (c). d) Density profile of the stationary state in semi-log scale. The exponential decay observed in the wing is the emblematic signature of Anderson Localisation~\cite{Anderson:1958}.  }
\label{fig:atomlaser}
\end{figure}

Since these early demonstration, new schemes have been developed. For instance, atom lasers can be produced by simply lowering the trap depth, which can be done in an ``all-optical" way\,\cite{Cennini:2003}. Such techniques do not require any transition between internal degrees of freedom and are intrinsically less sensitive to the surrounding magnetic field fluctuations, which constitute a major drawback of the ``genuine'' scheme. Besides, efforts have been made to achieve a better control on the beam propagation using new atom-optics tools, like reflectors\,\cite{Bongs1999}, or by compensating gravity\,\cite{Kleine:2010}. Atom lasers can also be directly coupled into waveguides, consisting in a laser guiding the atoms through the dipole force\,\cite{nous:GAL,Couvert:2008,Dall:2010}, similarly to the `pig-tailed' photonic lasers (see Fig. \ref{fig:atomlaser}). These configurations yield to nearly transverse single mode occupancy (see e.g.\,\cite{Gattobigio:2009PRA}  for a detailed analysis) and open the way to the realisation of integrated circuits\,\cite{Houde:2000,Torrontegui:2010}.

This short overview testifies to the interest dedicated to this new atomic source in the cold atom scientific community. Indeed, based on the example of its photonic counterparts, its coherence and brightness hold great promise to improve the sensitivity of atomic interferometers. Moreover, interactions between atoms can lead to interesting non linear atom optics phenomena: the use of solitons in atom interferometers\,\cite{Negretti2004}, or the reduction of atom number fluctuations with squeezed matter waves\,\cite{Louchet2010}. In addition, atom lasers constitute a very appropriate tool to probe fundamental concepts in physics, for instance the test of equivalence principle in general relativity or the study of a rich variety of quantum transport phenomena. Such prospects, discussed in the following, have triggered a significant experimental and theoretical effort to characterise the atom laser properties, including coherence, flux and spatial mode quality. 

First, the temporal coherence raises fundamental questions: will the well defined phase of the BEC be preserved? What about the phase diffusion along the propagation? Landmark experiments in the group of T. Esslinger have answered those questions by measuring the first\,\cite{Kohl:2001} and second\,\cite{Ottl:2005} order coherence of an atom laser. Provided that classical fluctuations have been suppressed, the coherence length was there shown to be Fourier limited by the outcoupling duration. Second, much work has been devoted to another key feature: the flux of the atom laser. Along these lines, quantitative descriptions of the outcoupling process have been made (see e.g.\,\cite{Gerbier:2001}) and a special attention has been paid to the weak-coupling conditions enabling a ``quiet'' and quasi-continuous emission\,\cite{Jack:1999,Moy:1999,Robins:2005}. These conditions imply quite stringent flux limitations that may constrain future applications. In fact these limits strongly depend on the specific atom laser scheme, Raman outcouplers allowing to reach the largest peak flux, typically a few 10$^9$ at.s$^{-1}$\,\cite{Robins:2006,Debs:2010}. Last, the transverse dynamics has been investigated, either in guided configurations (in term of transverse mode occupancy as mentioned above) or in the ``freely'' propagating case\,\cite{Busch:2002,Kohl:2005}. For the latter situation, the popular beam-quality factor $M^2$, initially introduced by Siegman\,\cite{Siegman:1991} for photon laser, has been adapted to describe the propagation of the atomic beam with ABCD matrices and to characterise how far an atom laser deviates from the diffraction limit\,\cite{nous:caustiques,Jeppesen:2008}. This work especially emphasises how atom optics can benefit from the methods originally developed in the optical domain.

More than a decade after its first realisation, the atom laser has now clearly gained in maturity and is heading towards applications. With the realisation of spatially separated beams\,\cite{Dugue:2008,Dall:2009}, progress has been made towards the integration of atom lasers into interferometric schemes, for instance to realise ultra-sensitive inertial sensors\,\cite{Cronin2009}. Moreover the atom laser pair created in\,\cite{Dall:2009} is expected to exhibit non classical correlations that could improve the interferometer sensitivity below the shot-noise limit\,\cite{Giovannetti:2004}. Another route towards sub-shot-noise interferometry is to take benefit from the inherent inter-atomic interactions. This non-linearity could be used to generate ``squeezed states'' along the propagation \,\cite{Johnsson:2007}, as the Kerr effect does with photons.

Besides atom interferometry, one of the major interest of ultracold atomic systems is their suitability as a model system to revisit fundamental concept in condensed matter physics\,\cite{Bloch2008RMP}. In this context, the horizontally guided atom lasers\,\cite{nous:GAL,Couvert:2008}, where the de Broglie wavelength is kept large and constant along the propagation, are particularly well suited to study quantum transport phenomena past obstacles (see e.g.~\cite{Paul:2007PRA}). A rich physics is here expected ranging from linear effects (e.g. tunnelling transmission, quantum reflection, Bloch oscillations in a periodic potential...) to nonlinear effects (for instance the atomic analogue of the Coulomb blockade through a micro-cavity). As we detail below, a striking example of the control achieved in these systems is given by the recent experiments studying the propagation through disorder. 

Considering transport properties, the role of disorder, whose presence cannot be avoided in a real material, comes naturally to mind. There, the subtle interplay between phase coherence, diffusion and inter-particle interactions leads to numerous complex phenomena which are not yet fully understood\,\cite{LSP:2009}. At the very heart of those, lies the emblematic Anderson localisation (AL), discovered in 1958\,\cite{Anderson:1958}. It predicts that even a small amount of disorder can completely freeze the motion of non-interacting particles, leading to a purely quantum metal-insulator transition. Past years, there has been a strong interest to directly observe AL with ultracold atoms. It finally succeeded at Institut d'Optique in Orsay\,\cite{Billy2008} (see Fig.~\ref{fig:atomlaser}), and simultaneously at LENS in Florence\,\cite{Roati2008} in quasi one-dimensional systems. Most importantly, these two landmark experiments are very promising in view of future extensions to more complex situations, i.e. in higher dimensions or with controlled interactions~\cite{AspectToday:2009,Robert:2010,Deissler:2010,Dries:2010, Pasienski2010}. Similar 1D AL was observed in momentum space in a cloud of thermal atoms\,\cite{Chabe2008}.

After these promising results, one may wonder if the atom laser will remain only a beautiful object of study for the physicist or will escape from the labs and lead to practical applications. The answer will likely depend on two main challenges: i) building miniaturised and reliable systems similar to the great technical improvement brought by semiconductor diode lasers  and ii) achieving continuous operation. As illustrated by the recent experiments realised in micro-gravity (either in a 146-meter-tall drop tower\,\cite{Zoest:2010} or in ``zero-g'' parabolic flights\,\cite{Stern:2009}), the first of those is about to be taken up. In parallel, important steps have been made towards continuous operation. Both the merging of two BECs\,\cite{Chikkatur:2002} and the simultaneous pumping and outcoupling\,\cite{Robins:2008} were indeed demonstrated. The combination with other approaches such as continuous condensation in an atomic beam\,\cite{Lahaye:2006,Olson:2006} or continuous loading of a trap\,\cite{Aghajani:2009} is then very promising.

\section{Optical dipole traps to quantum degeneracy}
\label{Laburthe}

The use of far detuned lasers to create optical dipole traps for cold atoms and reach quantum degeneracy has spread dramatically in the last ten years. While some of the reasons for this have been cited in the previous sections, the goal of this section is to highlight some of the technical advantages of optical dipole traps, and to discuss, briefly, how they open a whole new field of investigation related to quantum many-body physics.

\subsection{Figures of merit of the optical dipole trap}

Evaporation in magnetic traps has been the main route to the production of quantum degenerate gases, but the use of lasers to confine and evaporate atoms to quantum degeneracy is now taking over. To create a trap with cw lasers, one uses the inhomogeneous AC stark shift produced by tightly confining a red-detuned laser on the atoms\,\cite{Grimm2000}. Deep traps are typically created using very powerful lasers, and the first all-optical quantum gases were achieved using CO$_2$ cw lasers\,\cite{Barrett2001,Granade2002}. Cs could be first condensed using a combination of CO$_2$ and 1 $\mu$m lasers\,\cite{Weber2003}. With the increasing level of power that one can produce at 1\,$\mu$m in fibre lasers\,\cite{Kinoshita2005}, they have now become the lasers of choice for all optical BEC (due to more traditional optics, and smaller diffraction-limited spot size). A solid-state laser at 532~nm has been used to reach BEC for Yb\,\cite{Takasu2003}, and, more recently, cw fiber lasers at 1.5~$\mu$m have also been employed for Rb\,\cite{Clement2009}. This wavelength lies in the telecom domain, which allows to take advantage of important fibre-based technological developments.

The first burst of interest into all optical BEC was due to the fact that lasers provide a relatively fast way to produce BECs, as they typically produce traps with large oscillation frequencies. They may be more difficult to load than magnetic traps, due to a smaller volume, but once loaded, BECs can be produced in a matter of seconds, compared to typically 30 s in conventional magnetic traps.

Stability and flexibility are two additional benefits of using lasers to trap quantum degenerate gases. Stability is provided by the extremely low pointing and power noise of lasers; in addition, possible fluctuations in pointing or power can be compensated for by active stabilisation with a large bandwidth. The stability of laser traps is such that some experiments now start evaporation in magnetic traps (to take benefit of  their large volume) and finish evaporation in more stable optical traps\,\cite{Lin2009}. Using lasers to create traps also brings flexibility: typically, quantum degenerate gases are produced at the intersection of two focussed laser beams~\cite{Kuhn1997}, and the trap geometry can be controlled in real time by the ratio of power in each of the two beams. One can also use the periodic potential made by the interference of two or more coherent laser beams to trap atoms into optical lattices (as described in section~\ref{Lemonde} of this paper). Recently, holographic techniques were used to build traps of complex shape\,\cite{Newell2003,Bergamini2004}, and dark hollow traps based on Laguerre-Gauss profiles were demonstrated\,\cite{Olson2007}.

\subsection{What are dipole traps good for?}
Optical traps\,\cite{Grimm2000} rely on the AC stark shift, and therefore mostly depend on the electronic structure of the atom of interest. One can thus trap atoms with no magnetic moment (as is the case for Yb\,\cite{Takasu2003}) with interesting metrologic perspectives; another application is to trap atoms in their absolute ground state, in which, interestingly, two-body inelastic collisions at low temperature are energetically forbidden, but which, unfortunately, cannot be magnetically trapped.

When the laser frequency is very detuned compared to the resonant lines of the atom, the optical traps are almost independent of the Zeeman or hyperfine internal sub-state\,\cite{Grimm2000}. This has considerable impacts on the physics that one can explore with quantum degenerate gases: one can trap mixtures of atoms in different internal states with exactly the same trap, and explore new quantum phases involving the internal (spin) degrees of freedom. Collisions between non-polarised atoms are described by various scattering lengths, which provides spin-dependent contact interactions. This leads to spin dynamics (at constant magnetisation, since contact interactions are isotropic), and new quantum phases. For example, in the case of spin 1 bosons, the sign of spin exchange interaction determines which spin configuration is lowest in energy, and the ground state is either ferromagnetic or polar\,\cite{Ho1998,Machida1998}. The spinor phases  have been investigated for $F=1$ and $F=2$ atoms, using Rb and Na, by studying miscibility \cite{Stenger1998} and spin dynamics \cite{Chang2004,Schmaljohann2004,Kronjager2006,Black2007}. Interestingly, an interplay between the linear and the quadratic Zeeman effect can also be used to study quantum phase transitions between different multi-component BEC phases\,\cite{Sadler2006}. Finally, at very low magnetic fields and for mesoscopic samples, spinor gases should reach non-classical spin states \cite{Law1998} or lead to fragmented BECs \cite{Ho2000}.

At moderate detunings, the AC stark shift may also depend on the Zeeman substate (vectorial or even tensorial light shift)\,\cite{Grimm2000}. This has been used to control and address individual lattice sites in an array of double-wells using radio-frequency techniques\,\cite{Lee2007}, with interesting application to quantum computing\,\cite{Anderlini2007}. 

A last important interesting feature of optical dipole traps is that trapping is independent of the magnetic field. This is of utmost importance for the study of magnetically tunable Feshbach resonances\,\cite{Inouye1998}. In the case of mixtures of ultracold fermions, magnetically tunable Feshbach resonances in optical traps have led to the study of strongly interacting fermions, briefly described in the last paragraph of this section.

\subsection{Dipolar quantum gases}

Bose-Einstein condensates are quantum fluids, whose properties greatly depend on the interactions between particles. While most of the experiments performed up to now were in a regime where interactions are dominated by the short-range and isotropic van der Waals interactions, the production of Bose-Einstein condensates with highly magnetic Cr atoms \cite{Griesmaier2005,Griesmaier2006,Beaufils2008} allowed for the study of quantum gases where dipole-dipole interactions cannot be neglected. In contrast to the van-der-Waals interaction, dipole-dipole interactions are long-range and anisotropic, which introduces qualitatively new physics in the field of quantum degenerate gases\,\cite{Lahaye2009}.

Once Cr BECs are produced, as optical traps are insensitive to magnetic fields, one can use a magnetically tunable Feshbach resonance to decrease the scattering length of Cr atoms, hence increasing the relative strength of dipole-dipole interactions. It is then possible to produce an almost pure dipolar gas\,\cite{Lahaye2007}, which spontaneously collapses due to the attractive part of dipole-dipole interactions\,\cite{Lahaye2008}. During this collapse, the BEC loses its parabolic shape and reveals the symmetry of the dipolar interactions. The collapse dynamics depend on the trapping geometry\,\cite{Koch2008}, a direct consequence of the anisotropic character of dipole-dipole interactions. Dipolar interactions can also be revealed by studying collective excitations, which depend on the orientation of the dipoles relative to the trap axis. A departure from the usual behaviour of collective excitations of trapped Bose-Einstein condensates has recently been observed \,\cite{Bismut2010}. 

Dipolar interactions also open new perspectives in the field of spinor condensates, as they allow magnetisation changing collisions. For example, one expects that for some given experimental parameters, spin relaxation spontaneously induces vortices in Cr BECs, equivalently to the Einstein-de Haas effect\,\cite{Kawaguchi2006,Santos2006,Gawryluk2007}. It has been shown that magnetisation dynamics is modified in optical lattices, when the trapping frequency in each lattice site is on the order of the Larmor frequency\,\cite{Pasquiou2010,Pasquiou2011a}, a consequence of an inter-play between spin dynamics and physical rotation. Very recently, magnetisation dynamics due to dipole-dipole interactions was studied at extremely low magnetic fields, such that spin-dependent interactions dominate the linear Zeeman effect \,\cite{Pasquiou2011b}. 

The experimental activity on species with relatively important dipole-dipole interactions has been recently extended to Er and Dy atoms (up to now far from the quantum regime)\,\cite{Berglund2008,Lu2010}. Heteronuclear molecules can also possess a large induced electric dipole moment, leading to even stronger dipole dipole interactions, with interesting possible prospects for quantum computation, or the study of novel quantum phases (for example checkerboard or supersolid) \,\cite{Goral2002}. Progress towards this goal are impressive\,\cite{Ni2008}, but the experimental task remains challenging.

\subsection{Strongly interacting fermions}

The understanding of fermionic many-body systems is one of the most challenging problems in modern quantum physics: indeed, in addition to interactions, the antisymmetrisation of the wave-function imposed by Pauli Exclusion Principle implies the existence of non trivial quantum correlations making the resolution of the problem highly non trivial.

Combined to Feshbach resonances, the realisation of ultra-cold Fermi gases in dipole traps have paved the way to a new era in the experimental study of the quantum many body system by providing highly controllable systems corresponding to parameter ranges beyond the reach of usual condensed matter devices. The most striking example is the exploration of the connexion between Bardeen, Cooper and Schrieffer's theory (BCS) of superconductivity and Bose-Einstein condensation through the {\em BEC-BCS crossover} model. This theory, developed in the early eighties by Leggett\,\cite{leggett81}, Nozi\`eres and Schmitt-Rink\,\cite{nozieres1985bose}, addresses the properties of an attractive Fermi gas and shows that the BCS state describes the weakly attractive regime of large size Cooper pairs, while the strongly attractive regime is associated with the Bose-Einstein condensation of tightly bound dimers. Interest for this problem was revived in the 90's with the advent of high critical temperature superconductors: in these systems the size of the Cooper pairs is small compared to usual metallic superconductors, suggesting that they may operate in the transition region of the BEC-BCS crossover.
Using the ability to tune interactions using Feshbach resonances, cold atoms provided in 2003 the first experimental confirmation of the crossover scenario\,\cite{greiner2003emergence,jochim2003bose,Zwierlein03obe,bourdel2004esb}. The wide range of experimental investigation probes available in atomic physics has made possible an accurate quantitative description of the physical properties of the system, from the characterisation of superfluidity by the observation of quantized vortices\,\cite{zwierlein2005vortices}, the study of low-lying excitation modes\,\cite{bartenstein2004cmo,altmeyer2007PMC} or the measurement of the thermodynamic equation of state\,\cite{nascimbene2009eos,navon2010Ground}.

In more recent years, the physics of ultra-cold Fermi gases has evolved towards even more exciting new directions. It was first possible to explore the properties of spin-polarised fermionic systems \cite{zwierlein2006fsi,partridge2006pap,nascimbene2009pol}. These experiments have confirmed the Clogston-Chandrasekhar  scenario of resistance of Fermionic superfluidity against spin imbalance due to the presence of the gap characterising the superfluid state \cite{clogston1962ulc,chandrasekhar1962}. Theoretical \cite{pilati2008psp,parish2007finite} and experimental investigations have revealed a very rich phase diagram (see \cite{sheehy2010imbalanced,Chevy2010Unitary} for a review): In the BCS regime and near unitarity -- where the scattering length $a$ diverges -- the properties of the spin polarised system can be obtained from the properties of an impurity immersed in a Fermi sea, the so-called Fermi polaron \cite{lobo2006nsp,prokof'ev08fpb,chevy2006upa,combescot2007nsh}. The single particle properties  of the polaron have been characterised experimentally by radio-frequency spectroscopy \cite{schirotzek2009ofp} and the study of its eigenmodes \cite{nascimbene2009pol}. The theoretical \cite{lobo2006nsp,Mora2010Normal} and experimental \cite{navon2010Ground} study of the collective behaviour of this quasi-particle have shown that the gas of polarons could be described in the framework of Landau's Fermi liquid theory.  Up to now, only the normal polaron gas has been investigated experimentally but, based on a general argument by Kohn and Luttinger, it is expected that at low temperature, the gas of polarons can form a p-wave superfluid \cite{bulgac2006ipw}. In the BEC limit $k_{\rm F}a\lesssim 1$ (here $k_{\rm F}$ is the Fermi wave vector), the impurity turns from a fermionic to a bosonic behaviour. The system is then described as a mixture of bosonic point-like dimers immersed in the Fermi sea of excess atoms as predicted in  \cite{prokof'ev08fpb,mora2009ground,punk2009polaron,combescot2009analytical,Alzetto2010Equation} and reported experimentally in \cite{shin2008rsi}.

Magnetism in fermionic systems is another very active field of research, that may shed new light on long standing issues in solid state physics: recent experiments at MIT have raised the question of the existence of ferromagnetism in itinerant fermionic systems \cite{jo2009itinerant} (the Stoner instability \cite{stoner1933atomic}); in optical lattices, the groups of Zurich and Munich have investigated the Mott superfluid-insulator transition of repulsive fermions \cite{jordens2008mott,schneider2008metallic}. These breakthroughs open the way to the study of more exotic phases, such as antiferromagnetic order due to superexchange between neighbouring sites of the lattice, a phase that may be important in the understanding of high critical temperature superconductors.

Finally, mixtures of Fermi gases can provide a new and original test-ground for many body theories. Experiments have focused on $^6$Li/$^{40}$K mixtures for which several Feshbach resonances have been discovered and will allow for the exploration of strongly correlated regimes \cite{wille2008exploring,taglieber2008quantum,wu2011strongly,ridinger2011large}. The possibility of tailoring different optical potentials for the two atomic species broadens the range of phenomena accessible to experimental exploration. These novel systems can for instance be used to probe the Wigner crystallisation of dimers \cite{petrov2007crystalline}, Anderson localisation by atomic scatterers \cite{gavish2005matter} or even systems of mixed dimensionality where lithium atoms are free to move in three dimensions while potassium is confined in two dimensions, a system akin to brane physics in string theory \cite{nishida2008universal}.

\section{Toward laser cooling of molecules}
\label{Comparat}
\subsection{Introduction and state of the art}
Laser techniques such as precision spectroscopy or femtosecond control of  chemical reactions have improved considerably our knowledge on molecular physics. One of the greatest challenges of modern physical chemistry is to now push forward the techniques to probe and manipulate molecules in order to explore molecular physics at low temperature. 
A first ``cold regime'' is reached (for $T<1$ K) when the  de Broglie wavelengths become comparable to the size of molecules; then  resonances or tunneling effects dominate the chemistry. A second ``ultracold regime'' is reached (for $T<1$ mK), when s-wave scattering dominates, and fascinating collective effects eventually arise when the de Broglie wavelengths become comparable to the separation between molecules.

Unfortunately, the standard technique used for atoms to reach such low temperatures, namely laser cooling is generally not available for molecules because of the modification of the internal state occurring after the spontaneous emission step, which is the fundamental dissipative process in laser cooling.
Here, by \textit{cooling of molecules} \cite{1981PhRvL..46..236D} we mean the reduction of the translational temperature $T=T_{\rm trans}$ whereas the control of the rotational and the vibrational excitations ($T_{\rm rot}$ and $T_{\rm vib}$ temperatures being understood as best fitting parameters for an hypothetic  Boltzmann distribution), is called \textit{(optical) pumping of molecules}. 
 
Hence, the sub-Kelvin temperature ``cold molecule'' domain has been reached only in 1998 by associating cold atoms into molecules. Since then, many  other techniques have been developed either by starting from cold atoms or by manipulating already formed molecules as reviewed in  \cite{cold_Mol_2009}. In brief, the techniques starting with cold atoms associate them into molecular states by engineering a free-bound transition with a laser, a magnetic field, or by using collisional (three-body) processes. These methods have created ultra-cold molecules in the micro- or nano-kelvin temperature range and the quantum degenerate regime of a molecular Bose-Einstein condensate has been reached\cite{1998PhRvL..80.4402F,2003Sci...302.2101J,2003Natur.426..537G}. With few exceptions \cite{1999PhRvL..82..703N,2008PhRvL.101m3004D,2009PhRvA..79b1402V}, the molecules are usually in high vibrational states. On the other hand, the techniques starting with molecules, such as  cryogeny \cite{1998Natur.395..148W}, energy redistribution in supersonic beams \cite{1999PhRvL..83.1558B} or velocity filtering of effusive molecular beams \cite{2003PhRvA..67d3406R} are able to create molecules in low vibrational states but unfortunately with translational temperatures not colder than a few millikelvins. These techniques involve light forces, mechanical forces, as well as  the Stark or the Zeeman effects and apply to a large family of molecules. 

\begin{figure}[t]
\centering
\includegraphics*[width=0.75\textwidth]{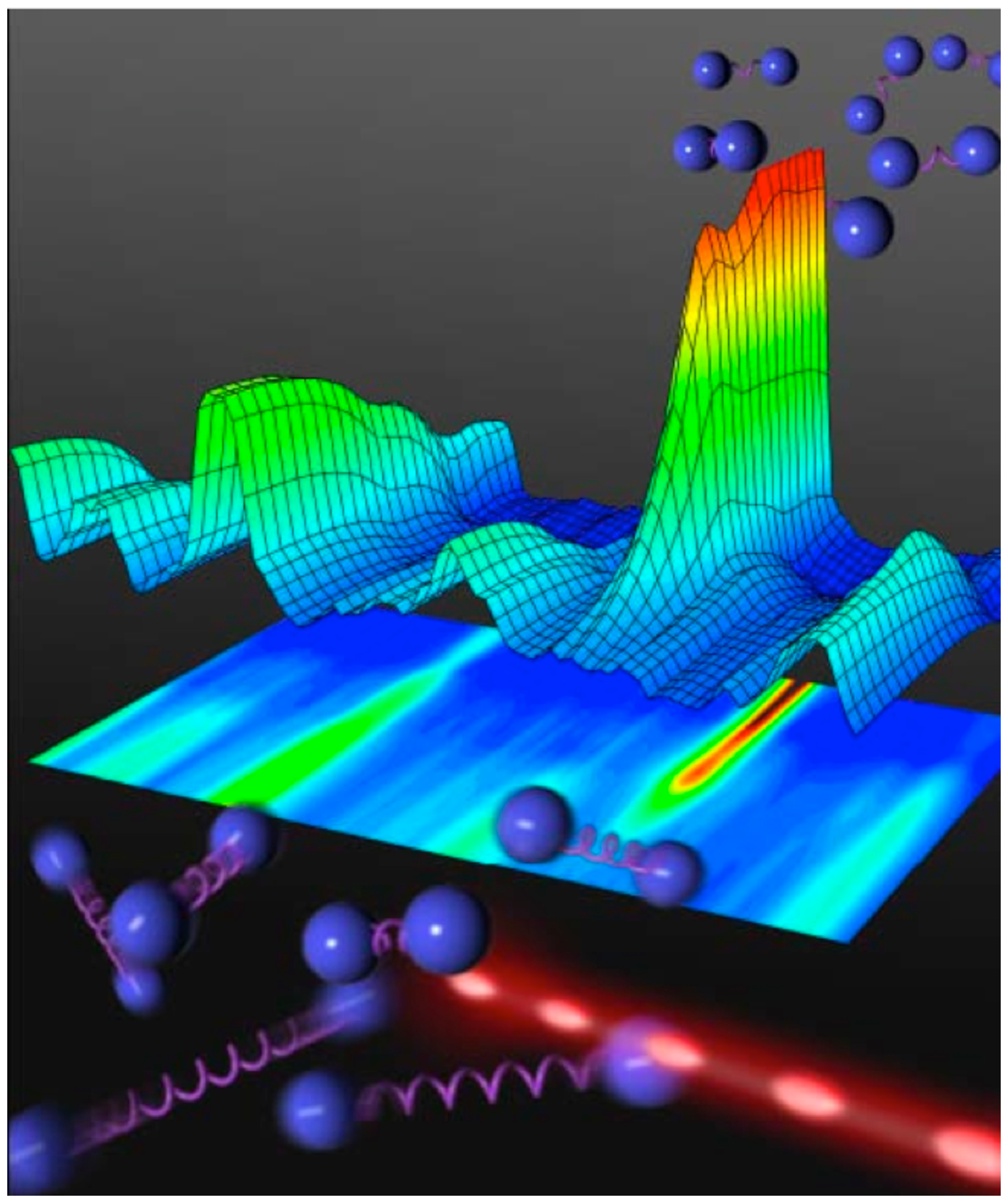}
\caption{Illustration of the optical pumping of molecules using train of shaped laser pulses. As times goes on (from bottom-left to up-right) the vibration of the molecules is modified by the laser pulses: all molecules end up in the same non vibrating state. The experimental signal, adapted from reference\,\cite{MatthieuViteau07112008}, shows the temporal evolution of the populations in the different vibrational levels of the ground state. The decreasing populations correspond to vibrational states $v=1-10$ whereas the increasing signal corresponds to the population of $v=0$.}
\label{fig:opt_pump}
\end{figure}

Thus, the cold molecule community has to face two major challenges. The first is to control the internal degree of freedom of molecules formed by atom-association. Impressive advances in lowering the internal state energy have been recently achieved either by transferring population of a single level 
 \cite{2005PhRvL..94t3001S,2008PhRvL.101m3005L,2010PhRvL.104c0402O,danzl2010ultracold}, with laser stimulated transitions  or by  a reduction of  $T_{\rm vib}$ through optical pumping of molecules using a spectrally shaped broadband laser as illustrated by Fig.~\ref{fig:opt_pump} \cite{MatthieuViteau07112008,2009PhRvA..80e1401S}. The second challenge is to lower the translational temperature of molecules, in analogy to atom cooling. A major breakthrough was obtained very recently by directly laser cooling a well suited molecule (SrF) with a quasi-closed-level system~\cite{Shuman2010}.
  
\subsection{Cooling schemes for molecules.}

The amount of possible cooling schemes for molecules, using optical, magnetic or electric forces combined with laser (or Radio-Frequency) transition of any kind is very large. Almost all the proposed ideas  have been suggested and often demonstrated for atoms. Here, we briefly describe a few of these schemes:

\paragraph{Collisional cooling.}

A first possibility is to use collisions with colder species such as trapped laser cooled atoms or ions in a so called sympathetic cooling scheme.
This technique has been demonstrated with molecular ions \cite{2000PhRvA..62a1401M}, but up to now reactive or inelastic collisions have ruled out the process for neutral species \cite{2009FaDi..142..191S}. As for other techniques such cooling methods may be improved by acquiring extra information on the sample, as in feedback or in stochastic cooling \cite{2001PhRvA..64f3410B}.
  
\paragraph{Laser cooling.}
		
Deflection of molecules, first observed in 1972 using electric field \cite{1972JChPh..57.1487D},  has been observed using laser radiation pressure as early as in 1979 \cite{herrmann1979molecular}. One could then expect that this would have opened the way to laser cooling. Only very recently this type of laser cooling of molecules has been observed with SrF~\cite{Shuman2010}. Unfortunately, laser cooling relies on several absorption-spontaneous emission cycles to remove kinetic energy, and closed-level schemes cannot generally be found for molecules.

However, it is probably possible to close the cycle for other molecules by using optical pumping as demonstrated in \cite{MatthieuViteau07112008,2009PhRvA..80e1401S}, or by using  an external
cavity \cite{2008PhRvA..77b3402L}. These techniques may enable laser cooling for a larger variety of molecules.

Finally, as pioneered  by Alfred Kaslter in 1950 \cite{Kastler1950}, another laser cooling process is possible in very dense gas, liquid or solid phases \cite{ISI:000264420800006}, taking benefit of collisions inducing thermal equilibrium  between $T_{\rm trans}$, $T_{\rm vib}$ and $T_{\rm rot}$. Indeed, such laser cooling of molecules has been realised in 1981 by removing vibrational energy of CO$_2$ molecules using a CO$_2$ laser \cite{1981PhRvL..46..236D}.
Unfortunately, the translational temperature reduction obtained was below the percent level. A more recent realisation on atoms shows a much higher efficiency\,\cite{Vogl2009}.

\paragraph{`Potential climbing' cooling.}

The last, and maybe the most promising and general idea is to remove kinetic energy by transferring it into potential energy using external forces, and using (absorption-)spontaneous emission to make the process irreversible \cite{Ovchinnikov1997,2009NJPh...11e5046N,2008PhRvL.100x0407T,Falkenau2011}. Such one-way cooling can even be repeated, in a Sisyphus-like process \cite{1985JOSAB...2.1707D,1994OptCo.106..202H}, for instance by bringing back the particles  to their original state \cite{2009PhRvA..80d1401Z,2009NJPh...11e5046N,2009JPhB...42s5301R}. 
	
%\end{enumerate}

A widespread method  to realise large sample of molecules at very low temperatures has still to be demonstrated. Laser techniques combined with external potentials will probably play a key role in this quest. The challenge is nevertheless appealing because molecules at low temperatures are expected to lead to significant advances in laser molecular spectroscopy, optical molecular clocks, fundamental tests in physics, controlled laser-chemistry studies and quantum computation thought control of quantum phenomena  \cite{cold_Mol_2009}.

\section{Conclusion}
In this review, we have discussed some of the recent advances on the physics of ultracold atoms and molecules allowed by the advent of lasers. It was not possible in the limited length of this paper to comment on all the aspects of this fast developing research area. In particular, the great progress in the applications of laser cooling and laser control to ions for metrology or quantum information\,\cite{Wineland2011} was not discussed here.

The first field which benefited directly from the progress in laser cooling and trapping is atom interferometry and especially time and frequency metrology. Cold caesium fountain clocks have become the regular way of building a primary frequency standard. Recent progress in optical clocks confirm the importance of laser manipulation of atoms in the development of new frequency standards. This allows both dramatic improvements in inertial sensors and time metrology, and opens the way to new tests of fundamental physics -- variation of the fundamental constants or test of general relativity\,\cite{Chou2010b,Flambaum2007}.

After 15 years of Bose-Einstein condensation in dilute gases, the field has become mature and looks for applications in metrology\,\cite{Hughes2009,Gross2010} and condensed matter. Experiments where cold atoms mimic difficult condensed matter problems are now within reach. Both periodic systems confined in optical lattices and low dimensional systems have shown properties specific to strongly correlated systems\,\cite{Bloch2008RMP}. When considering the internal (spin) degree of freedom, the nature of the ground state depends on the relative interactions. Both ferromagnetic and polar behaviour have been evidenced, as well as quantum phases transitions\,\cite{Sadler2006}, and spin textures. The amazing control over important parameters such as the atomic density, the interactions or the temperature opens the way to the metrology of many-body bosonic or fermionic systems, as well as the study of few body physics\,\cite{Kraemer2006}.

Whereas dramatic achievements have been accomplished, important challenges remain. First, only one fourth of the stable atomic species, and a few tens of molecules, are available at low temperatures. Cooling different atomic species is technologically involved due to the complex internal atomic or molecular structure. The recent laser cooling of SrF\,\cite{Shuman2010} is promising in this direction, which may eventually lead to cold (quantum) chemistry. Second, the coupling of degenerate gases with external devices like high finesse cavities\,\cite{Brennecke2008}, superconductors\,\cite{Roux2008,Hufnagel2009} or nano-objects\,\cite{Vetsch2010} has just started. Third, a better control on the temperature will be necessary for the implementation of a quantum simulator for condensed matter problem, for example to determine relevant phase diagrams near zero temperature. Finally, the control of entanglement in a many body system is a key feature both for the study of many body physics itself and for quantum computation. Recently, entanglement found an application as a new diagnostic tool in metrology: quantum logic spectroscopy, based on the entanglement between two ions in a common trap, recently allowed unprecedented accuracy in Al$^+$ clocks\,\cite{Chou2010a}.

\section*{Acknowledgements}
We thank Mich\`ele Leduc for giving us the opportunity to write this review in the name of IFRAF. All the authors are members of the Institut Francilien de Recherche sur les Atomes Froids (IFRAF).

\bibliographystyle{CRASLN}
\bibliography{Bib_vinz,bib_phil,CRAS,2010_bibli_global_juin,bibliographie}

\end{document}